\documentclass[11pt]{article}
\usepackage{graphicx}
\usepackage{amssymb}
\usepackage{amsmath}
\usepackage{graphicx}
\usepackage{amstext}
\usepackage{esint}

\def\beq{\begin{eqnarray}}    
\def\eeq{\end{eqnarray}}      

\newcommand{\OMo}{\Omega_{M}^0}

\newcommand{\OLo}{\Omega_{\Lambda}^0}

\newcommand{\rmo}{\rho_{m}^0}

\newcommand{\rM}{\rho_M}
\newcommand{\rmr}{\rho_m}

\newcommand{\wm}{\omega_m}

\newcommand{\rL}{\rho_{\CC}}

\newcommand{\rLo}{\rho_{\CC}^0}

\newcommand{\CC}{\Lambda}

\newcommand{\nueff}{\nu_{\rm eff}}

\newcommand{\LQCD}{\Lambda_{\rm QCD}}
\newcommand{\OMB}{\Omega_B}
\newcommand{\OMBo}{\Omega^0_B}
\newcommand{\ODM}{\Omega_{\rm DM}}
\newcommand{\ODMo}{\Omega^0_{\rm DM}}

\newcommand{\be}{\begin{equation}}
\newcommand{\ee}{\end{equation}}

\newcommand{\mysection}[1]{\section{#1}
\renewcommand{\theequation}{\thesection.\arabic{equation}}
\setcounter{equation}{0}}

\begin{document}



 \hyphenation{cos-mo-lo-gi-cal
sig-ni-fi-cant}




\begin{center}
{\large \textsc{\bf DARK MATTER, DARK ENERGY AND THE TIME EVOLUTION OF MASSES IN THE UNIVERSE}} \vskip 2mm

 \vskip 8mm

\textbf{Joan Sol\`{a}$^{1,2}$}

\vskip0.5cm

$^{1}$ High Energy Physics Group, Dept. ECM, Univ. de Barcelona,\\
Av. Diagonal 647, E-08028 Barcelona, Catalonia, Spain

\vskip0.5cm

$^{2}$  Institut de Ci{\`e}ncies del Cosmos\\
Univ. de Barcelona, Av. Diagonal 647, E-08028 Barcelona

\vskip0.5cm

E-mail:  sola@ecm.ub.edu
 \vskip2mm

\end{center}
\vskip 15mm

\begin{quotation}
\noindent {\large\it \underline{Abstract}}. The traditional ``explanation'' for the observed acceleration of the
universe is the existence of a positive cosmological constant.
However, this can hardly be a truly convincing explanation, as an
expanding universe is not expected to have a static vacuum energy
density. So, it must be an approximation. This reminds us of the
so-called fundamental ``constants'' of nature. Recent and past
measurements  of the fine structure constant and of the
proton-electron mass ratio suggest that basic quantities of the
standard model, such as the QCD scale parameter $\LQCD$, might not
be conserved in the course of the cosmological evolution. The masses
of the nucleons and of the atomic nuclei would be time-evolving.
This can be consistent with General Relativity provided the vacuum
energy itself is a dynamical quantity.  Another framework realizing
this possibility is QHD (Quantum Haplodynamics), a fundamental
theory of bound states. If one assumes that its running couplings
unify at the Planck scale and that such scale changes slowly with
cosmic time, the masses of the nucleons and of the DM particles,
including the cosmological term, will evolve with time. This could
explain the dark energy of the universe.
\end{quotation}

\vskip 8mm



\newpage

\vskip 6mm

 \noindent \mysection{Introduction}
 \label{Introduction}



There is no doubt that the origin and nature of the dark energy in
our universe is one of the deepest mysteries we can think of in
theoretical particle physics and cosmology. The bare fact is that
our universe is in an state of accelerated expansion and we have to
find an explanation for it. While the traditional ``explanation'' is
the existence of a nonvanishing and positive cosmological constant,
$\CC$,  whose energy-density equivalent $\rL=\CC/8\pi G$ is of order of the
critical density, this cannot be a truly convincing explanation, as
an expanding universe is not expected to have a static vacuum energy
density. Ultimately this is the main difficulty behind the so-called
cosmological constant problem\,\cite{CCproblem} in the context of quantum field theory (QFT) in curved spacetime\,\footnote{See e.g.
\cite{JSP-CCReview2013,SaintPetersburg2013} for recent reviews on
the role played by the dynamical vacuum energy in cosmological evolution.} .

The CC problem is the main source of headache for every theoretical
cosmologist confronting his/her predictions with the observational
value of $\rL$\,\cite{CosmoObservations}. After the discovery of a
Higgs-like boson at the LHC and the absence of new physics, many question marks are left open\,\cite{Altarelli2014}. The CC problem is actually the most severe one. In point of fact, it became even more acute than before since it is  reinforced by the fact that there is indeed a big vacuum
contribution generated in the SM which is triggered by the
phenomenon of spontaneous symmetry breaking (SSB) in the electroweak
sector of the  model. It is therefore more pressing than ever to
properly address the notion of vacuum energy and its possible
implications in cosmology\,\cite{JSP-CCReview2013}. Let us, however, not underemphasize the fact that to achieve such aim one has to face nontrivial problems of QFT in curved spacetime\,\cite{SS0809}.

In the last years an independent source of puzzling news has
generated also a lot of interest. Frequent hints that  the
electromagnetic fine structure constant $\alpha_{\rm em}$ might
change with the cosmic evolution are reported in the
literature\,\cite{MurphyWebbFlaumbaum2003} -- for reviews, see e.g. \cite{FundamentalConstants}. It is tempting to think
that if $\alpha_{\rm em}$ evolves with time, all of the fundamental
coupling constants should change in time, including the gravity
constant\,\cite{FritzschSola2012,FritzschSola2014}, see also \cite{Ward2014}. Since the
gravity constant $G_N$ determines the Planck mass $M_P=G_N^{-1/2}$,
one expects that $M_P$ depends also on time and slowly evolves with
the cosmic expansion. Planck satellite data is also sensitive to this kind of subtle effects on the fundamental ``constants''\,\cite{PLanck VariationFundConst}.

Such framework obviously implies a link between gravity and particle
physics. Here we wish to signal some possible connections and at the
same time describe particular theoretical frameworks where these
ideas could be implemented.


\mysection{Dynamical vacuum models}
\label{sect:Dynamical vacuum}

In a cosmological context with dynamical parameters $\CC$ and $G_N$
it is useful to consider the possible modifications that may undergo
the basic conservation laws.  The Bianchi identity satisfied by the
Einstein tensor on the \textit{l.h.s.} of Einstein's equations reads
$\nabla^{\mu}G_{\mu\nu}=0$, where $G_{\mu\nu}=R_{\mu\nu}-(1/2)
g_{\mu\nu} R$. It follows that the covariant derivative of the
\textit{r.h.s.} of Einstein's equations must be zero as well:
$\bigtriangledown^{\mu}\,\left(G_N\,\tilde{T}_{\mu\nu}\right)=0$,
where $\tilde{T}_{\mu\nu}\equiv T_{\mu\nu}+g_{\mu\nu}\,\rL $ is the
full energy-momentum tensor of the cosmic fluid composed of matter
and vacuum. Using the explicit form of the
Friedmann-Lema\^{\i}tre-Robertson-Walker (FLRW) metric, the
generalized conservation law emerging from this dynamical framework
reads:
\begin{equation}\label{BianchiGeneral}
\frac{d}{dt}\,\left[G_N(\rmr+\rL)\right]+3\,G_N\,H\,(1+\wm)\rmr=0\,,
\end{equation}
where $\wm=p_m/\rmr$ is the equation of state (EoS) for matter.
Consider now the following scenarios:

Scenario I: $\rL=\rL(t)$ is assumed variable, and $G_N=$const. In
this case, Eq.\,(\ref{BianchiGeneral}) implies
\begin{equation}
\dot{\rho}_{m}+3(1+\omega_{m})H\rho_{m}=-\dot{\rho}_{\Lambda}\,.
\label{frie33}
\end{equation}
Since in this case $\dot{\rho}_{\Lambda}\neq 0$ it means we permit
some energy exchange between matter and vacuum, e.g. through vacuum
decay into matter, or vice versa. Obviously, if
$\dot{\rho}_{\Lambda}=0$ we recover the standard covariant
matter conservation law:
$\dot{\rho}_m+3\,H\,(1+\wm)\rmr=0$. Its solution in terms of the scale factor
is well-known:
\begin{equation}\label{solstandardconserv}
\rmr(a)=\rmo\,a^{-3(1+\wm)}\,.
\end{equation}

Scenario II: $\rL=\rL(t)$ is again variable, but $G_N=G_N(t)$ is
also variable. In contrast to the previous case, here we further assume
matter conservation in the standard form (\ref{solstandardconserv}).
As a result the following conservation law ensues:
\begin{equation}\label{Bianchi1}
(\rmr+\rL)\dot{G}_N+G_N\dot{\rho}_{\CC}=0\,.
\end{equation}
In this case the evolution of the vacuum energy density is possible
at the expense of a running gravitational coupling: $\dot{G}\neq 0$.

Scenario III: Suppose we keep $\rL=$const., but $G_N=G_N(t)$ is again
variable. Now we find:
\begin{equation}\label{dGneqo}
\dot{G}_N(\rmr+\rL)+G_N[\dot{\rho}_m+3H(1+\wm)\rmr]=0\,.
\end{equation}
Here matter is again non-conserved and the gravitational coupling is
running. Despite the vacuum energy is constant in this scenario,
such situation can mimic a form of dynamical dark energy since it
implies a different expansion rate, a fact that could be detected through the effective equation of state of the dynamical dark energy that it gives rise to\,\cite{BasSola2013}.

The above three generalized cosmological scenarios differ from the
concordance $\CC$CDM model, but can stay sufficiently close to it if
we consider the recent history of our universe. Let us finally note that the above dynamical vacuum models can be appropriately extended at high energies for a successful explanation of the inflationary universe through the primeval vacuum decay\,\cite{H4model}.

\mysection{Fundamental constants and their possible time variation}
\label{sect:Fundamental Constants}

It has been proposed in \cite{FritzschSola2012} that he cosmic time
variation of $\CC$ and $G_N$ could be related to that of particle
masses. This is a challenging possibility. Let us take for instance
the proton mass whose current value is $m_p^0=938.272 013(23)$ MeV.
It can be computed from QCD using the scale parameter $\LQCD={\cal
O}(200)$ MeV, the quarks masses and the electromagnetic
contribution:
\begin{eqnarray}\label{eq:ProtonMass}
m_p &=&c_{\rm QCD}\LQCD+c_u\,m_u+c_d\,m_d+c_s\,m_s+c_{\rm
em}\LQCD\,,
\end{eqnarray}
where the bulk of the contribution ($860$ MeV) comes from the first
$\LQCD$ term on its \textit{r.h.s.}. Recall that the QCD scale
parameter is related to the strong coupling constant
$\alpha_s=g_s^2/(4\pi)$. To lowest (1-loop) order one finds:
\begin{equation}\label{alphasLQCD}
\alpha_s(\mu)=\frac{2\pi}{{b}\,\ln{\left(\mu/\LQCD\right)}}\,.
\end{equation}
Here $b=11-2n_f/3$ is the one-loop $\beta$-function coefficient,
with $n_f$ the number of quark flavors, and $\mu$ is the
renormalization point.

The value of $\LQCD$ could change with the cosmic expansion, and
thus be a function of the Hubble function $H$. In this case
$\alpha_s(\mu; H)$ would run both with the renormalization scale
$\mu$ and the Hubble funtion, which has also natural dimension of
energy. One can easily show that the relative cosmic variations of
the two QCD quantities are related (at one-loop) by:
\begin{equation}\label{eq:timealphaLQC}
\frac{1}{\alpha_s}\frac{d\alpha_s(\mu;H)}{dH}=\frac{1}{\ln{\left(\mu/\LQCD\right)}}\,\left[\frac{1}{\LQCD}\,\frac{d{\Lambda}_{\rm
QCD}(H)}{dH}\right]\,.
\end{equation}
If the QCD coupling constant $\alpha_s$ or the QCD scale parameter
$\LQCD$ undergo a small cosmological time shift, the nucleon masses
and the masses of the atomic nuclei would change accordingly.

Let us note on general grounds that as soon as one assumes
that the electromagnetic fine structure constant $\alpha_{\rm
em}$ can be varying, one expects the masses of all nucleons
to vary as well, since the interaction responsible for the variation of $\alpha_{\rm em}$ should couple radiatively to nucleons. In this sense one also expects the proton and neutron masses to be time dependent\,\cite{OlivePospelov 2002}.

Another clue to the time variation of masses is the following. In a grand unified theory (GUT) the various gauge couplings converge
at the unification point, and we can assume that they display the
double running form $\alpha_i=\alpha_i(\mu;H)$. One can show that
the GUT condition links the cosmic running of the electromagnetic fine structure constant $\alpha_{\rm
em}(\mu;H)$  to that of $\LQCD(H)$. It turns out that, under these conditions, $\LQCD$ runs
$\sim 30$ times faster than the electromagnetic fine structure
constant\cite{FritzschSola2012,FritzschCalmet}. Searching for a cosmic evolution of
$\LQCD$ is therefore much easier than searching for the time
variation of $\alpha_{\rm em}$!

\mysection{Cosmic acceleration versus time evolving masses} \label{sect:Cosmic acceleration and masses}

The different classes of cosmological scenarios considered in
Sect.\,1  could help us to understand the potential cosmic time
variation of the fundamental ``constants'' of nuclear and particle
physics, such as the QCD scale, the nucleon mass and the masses of
nuclei.

A class of dynamical vacuum models can be singled out. If the vacuum
energy density evolves as a function of the
form\,\cite{JSP-CCReview2013,SaintPetersburg2013}
\begin{equation}\label{eq: GeneralRG}
\rL(t)=c_0+\sum_{k} \alpha_{k} H^{2k}(t)+\sum_{k}
\beta_{k}\dot{H}^{k}(t)\,,
\end{equation}
with $c_0\neq 0$ (viz. an ``affine'' function constructed out of
powers of $H^2=\left(\dot{a}/a\right)^2$ and
$\dot{H}=\ddot{a}/a-H^2$, hence with an even number of time
derivatives of the scale factor $a$), one can formulate a unified
model of the cosmological evolution, compatible with the general
covariance of the effective action, in which inflation is predicted,
a correct transition (``graceful exit'') into a radiation phase can
be naturally accommodated, and finally the late time cosmic
evolution can also be successfully described.  For simplicity we
assume $\beta_k=0$. Furthermore, for the low-energy universe it
suffices to take the single term $k=1$ in (\ref{eq: GeneralRG}).
Therefore we are left with the simplest and yet nontrivial model
\begin{equation}\label{QFTModelLowEnergy}
\rho_{\Lambda}(H)=\rLo+\frac{3\,\nu}{8\pi}\,M_P^2\,(H^2-H_0^2)\,,
\end{equation}
where we have normalized such that $\rL(H_0)=\rLo$ is the current
vacuum energy density. We have also introduced the dimensionless
coefficient $\nu$ which we expect $|\nu|\ll 1$ such that the model
(\ref{QFTModelLowEnergy}) remains very close to the $\CC$CDM one --
see \cite{JSP-CCReview2013} for further details. One finds
$|\nu|={\cal O}(10^{-3})$ when confronting the model with
observations on type Ia supernovae, the Cosmic
Microwave Background, the Baryonic Acoustic Oscillations and structure formation\,\cite{BPS09}, a result which is compatible with the recent limits reported by the Planck satellite on the possible variation of the fundamental constants\,\cite{PLanck VariationFundConst}.

From (\ref{frie33}),(\ref{QFTModelLowEnergy}) and using Friedmann's
equation, one can solve for the matter and vacuum energy densities
as a function of the redshift in the matter-dominated epoch
($\omega_m=0$):
\begin{equation}\label{mRGa1}
\rho_m(z;\nu) =\rho_m^0\,(1+z)^{3(1-\nu)}\,,
\end{equation}
and the vacuum energy density:
\begin{equation}\label{CRGa1}
\rL(z;\nu)=\rLo+\frac{\nu\,\rho_m^0}{1-\nu}\,\left[(1+z)^{3(1-\nu)}-1\right]\,.
\end{equation}
For $\nu=0$ the matter density reduces to
Eq.\,(\ref{solstandardconserv}), and the vacuum energy density stays
constant at $\rLo$, as expected. From these expressions we can
determine the relative time variation of the matter nonconservation.
Define $\delta\rho_M\equiv \rM(z;\nu)-\rM(z;\nu=0)$ as the net
amount of non-conservation of matter per unit volume at a given
redshift $z$ and let us indicate by a dot the time variation. For
small redshifts, we find:
\begin{equation}\label{eq:deltadotrho2}
 \frac{\delta\dot{\rho}_M}{\rM}\simeq 3\nu\,\,H\,.
\end{equation}
and ${\dot{\rho}_{\CC}}/{\rL}\simeq -3\nu\,\,({\OMo}/{\OLo})\,H$.
The variation of the vacuum energy (compensating for the amount of
nonconservation of matter) is of opposite sign, as expected.

Let us be more precise. Take e.g. the baryonic density in the
universe, which is essentially the mass density of protons. We can
write $\rho_p=n_p\,m_p$, where $n_p$ is the number density of
protons and $m_p^0$ is the current proton mass. If the mass density
is non-conserved, it may be due to the fact that the proton mass
$m_p$ does not stay strictly constant with time and scales mildly
with the cosmic evolution:
\begin{equation}\label{eq:nonconservedmp}
m_p(a)=m_p^0\,a^{3\nu}\,, \ \ \ \ \ (|\nu|\ll 1)\,.
\end{equation}
Combining this equation with $n_p=n_p^0\,a^{-3}$ (the normal
particle number dilution law associated to the cosmic expansion,
with $n_p^0$ the current number density of protons) we find that the
proton density at any time is $\rho_p=\,n_p\,m_p=\rho_p^0
a^{-3(1-\nu)}$.

The index $\nu$ above could have been called $\nu_B$ since it affects the non-conservation of the baryon masses, such as the proton mass. In addition, being the matter content of the universe dominated by
the dark matter, we cannot exclude that these particles also vary
with cosmic time in a similar way, although perhaps with a different
anomaly index $|\nu_X|\ll 1$, such that
\begin{equation}\label{eq:nonconservedmX}
m_X(a)=m_X^0\,a^{3\nu_X}\, \,, \ \ \ \ \ (|\nu_X|\ll 1)\,.
\end{equation}
It is important to emphasize that since there is no a priori reason for the baryons and dark matter particles to follow the same rate of mass non-conservation, we may assume $\nu_X\neq\nu_B$ and hence we do not expect, in general, that this theory can just be rewritten as a mere $G$-varying theory, i.e., as a scalar-tensor theory.

From the above equations we can derive the time volution of the QCD
scale, and then from (\ref{eq:ProtonMass}) the time evolution of the
nucleon masses. We consider the total matter density of the universe
as the sum of nucleons and DM particles, but to simplify the
analysis we assume that $\nu_X=0$. In this case, introducing
$\nueff={\nu}/(1-\OMB/\ODM)$, we find:
\begin{equation}\label{eq:LQCDH}
\LQCD(H)=\LQCD^0\,\left[\frac{1-\nu}{\OMo}\,\frac{H^2}{H_0^2}-\frac{\OLo-\nu}{\OMo}\right]^{-(\ODMo/\OMBo)\,\nueff/(1-\nu)}\,,
\end{equation}
with $\OMo=\OMBo+\ODMo$. With this equation we can e.g. use
Ref.\,\cite{Reinhold06} comparing the $H_2$ spectral Lyman and
Werner lines observed in the Q 0347-383 and Q 0405-443 quasar
absorption systems with the corresponding spectral lines at present.
It involves redshifts in the range $z\simeq 2.6-3.0$ corresponding
to $12$ billion years ago. Assuming that $|\nu|={\cal O}(10^{-3})$,
as indicated before, we find that the relative variation of $\LQCD$
in this lengthy time interval is only at the few percent level with
respect to its present day value. It is, however, sufficient to be
sensitive to the most modern measurements planned in the near
future\,\cite{FritzschSola2012}.

\mysection{QHD: a fundamental theory of bound states} \label{sect:QHD}

Quite likely the standard model (SM) of strong and electroweak
interactions is not the final theory of the universe. The dark
matter (DM) and dark energy (DE) are fundamental problems awaiting
for an explanation. As an illustration of the kind of conceptual
modification that may be necessary to solve these problems, let us
consider the possible impact of Quantum Haplodynamics
(QHD)\,\cite{Fritzsch2012}.

In QHD all of the SM particles (except the photon and the gluons)
are bound states of the fundamental constituents called ``haplons'',
$h$, and their antiparticles. The idea was first formulated long ago
-- see \cite{FritzschMandelbaum,Barbieri}. Following
\cite{FritzschSola2014} we extend it and assume that the QHD chiral
gauge group is the unitary left-right group $SU(2)_L\times SU(2)_R$,
which we will denote $SU(2)_h$ for short.  All species of haplons
$h$ are $SU(2)$ doublets, hence each one has two internal states
$h_i$ represented by the $SU(2)_h$ quantum number $i=1,2\,$.
Rotations among these states are performed by the exchange of two
sets of massless $SU(2)_h$ gauge bosons
$\left(X_{L,R}^r\right)_{\mu}\,(r=1,2,3)$ for each chirality. There
are six haplon flavors, two of them are electrically charged chiral
spinors ($\chi=\alpha,\beta$) and four are charged scalars $S$. One
scalar ($\ell$) has electric charge (+1/2) and carries leptonic
flavor. The other three scalars have charge (-1/6) and carry color:
$c_k=R,G ,B$ (``red, green, blue''). In Table 1 we indicate
the relevant quantum numbers.

\begin{table}[t]
\begin{center}
\begin{tabular} {|c|c|c|c|c|}
  \hline
  $ \phantom{x}$ & $s$ & $Q$  & $SU(3)_c$ & $SU(2)_h$ \\ \hline\hline
  $\alpha$\phantom{x} & $1/2$ & $+1/2$ & $1$ & $2$ \\
  $\beta$\phantom{x} & $1/2$ & $-1/2$ & $1$ & $2$ \\
  $\ell$\phantom{x} & $0$ & $+1/2$ & $1$ & $2$ \\
  $c_k$\phantom{x}  &  $0$ & $-1/6$ & $3$ & $2$ \\
  \hline
\end{tabular}
\caption[]{Quantum numbers of the six haplons: spin ($s$), electric
charge $Q$ (in units of $|e|$) and corresponding representations of
$SU(3)_c$ and $SU(2)_h$.}
\end{center}
\end{table}

From the various haplon flavors the bound states of QHD can be
constructed. Only for energies $\mu$ well above  $\Lambda_h$  these
states break down into the fundamental haplons.  The weak gauge
bosons are s-wave bound states of left-handed haplons $\alpha$ or
$\beta$ and their antiparticles: $W^+  = \bar{\beta} \alpha$, $W^-
=  \bar{\alpha} \beta$ and $W^3  =\left(\bar{\alpha} \alpha -
\bar{\beta} \beta \right)/\sqrt{2}$.

The neutral weak boson mixes with the photon (similar to the mixing
between the photon and the neutral $\rho$-meson. One obtains the
physical $Z$-boson with a mass slightly heavier than the $W$-boson.
The LH confinement scale $\Lambda_h^L$ for $SU(2)_L$ defines  the
Fermi scale $G_F^{-1/2}\sim 0.3$ TeV and the size of the weak gauge
bosons of the SM. The confinement scale $\Lambda_h^R$ for $SU(2)_R$
is much larger (in the few TeV range) and has not been observed yet.

The leptons and quarks are themselves bound states. They are
composed of a chiral haplon ($\alpha$ or $\beta$) and a scalar
haplon: $\ell$ for leptons and $c_k$ for quarks. The electron and
its neutrino have the structure ${\nu} =({\alpha}\bar{\ell})$ and
$e^-=({\beta}\bar{\ell})$, which is consistent with the quantum
numbers of Table I. Similarly, the up and down quarks (with $c_k$
color) are given by: $u =({\alpha}\bar{c}_k)$ and $d
=({\beta}\bar{c}_k)$. Among the observed states, one of them has
zero haplon number and could be the resonance observed at the
LHC\,\cite{Fritzsch2012}. The outcome is an effective theory
equivalent to the electroweak SM in good approximation.

Additional particles are also predicted in QHD. The simplest neutral
bound state of the four scalars with haplon number ${\cal H}=4$ is a
stable color singlet spinless boson: $D =(lRGB)$. It is stable due
to haplon number conservation, which is similar to the conservation
of baryon number. Its mass is expected to be in the region of
several TeV. It can be produced together with its antiparticle by
the LHC-accelerator, and it can be observed by the large missing
energy. We interpret it as the particle providing the DM in the
universe. The properties of this DM particle are similar to a
``Weakly Interacting Massive Particle'' (WIMP), but it can be much
more elusive concerning the interactions with nuclei.

We can estimate the cross section for the D-boson off a nucleon
${\cal N}$ (of mass $m_{\cal N}$) as follows:
\begin{equation}\label{QHDcross-section}
\sigma_{\scriptsize D{\cal N}}\sim
f_D^2\,\frac{\alpha_h^2}{\left(\Lambda_h^L\right)^4}\,m_{\cal
N}^2\sim f_D^2\,\alpha_h^2\,G_F^2\,m_{\cal N}^2\,,
\end{equation}
where $G_F^{-1/2}\sim\Lambda_h^L\sim 300$ GeV  according to our
definition of Fermi's scale in QHD. Here $f_D$ is the dimensionless
form factor of the $D$-meson, which describes the confinement of the
haplons by the $SU(2)_h$ strong gauge force. All QHD bound states
have a form factor, which is of order one only for gauge boson
mediated interactions, which are described by the exchange of weak
bosons ($M_W^2\lesssim\left(\Lambda_h^L\right)^2$). For a deeply
bound state as $D$, however, we rather expect
$f_D\sim{\left(\Lambda_h^L\right)^2}/{B_D^2}\ll 1$, where $B_D$ is
the characteristic binding energy scale. For $B_D\simeq 5-10$ TeV
the scattering cross-section of $D$-bosons off nucleons,
Eq.\,(\ref{QHDcross-section}), can be reduced to the level of
$\lesssim10^{-45}$ cm$^2$, which is compatible with the current
stringent bounds\,\cite{LUXbounds}.

\mysection{Unification at the Planck Scale} \label{sect:Unification MP}

The QHD, QCD and QED  couplings might unify at the Planck scale. It
could have nontrivial implications for a possible explanation of the
DE in the universe, as we shall see. We can verify this possible
unification at one-loop level, starting from their low-energy values
and using the renormalization group equations (RGE's) to compute the
running of these parameters. For $SU(N)$ groups ($N>1$) one has:
\begin{equation}\label{RGE}
\frac{d\alpha_i}{d\ln\mu}=
-\frac{1}{2\pi}\left(\frac{11}{3}\,N-\frac23\,n_f-\frac16\,n_s\right)\,\alpha_i^2\equiv
-\frac{1}{2\pi}\,b_N\,\alpha_i^2\,,
\end{equation}
Here we have $\alpha_i=\alpha_h,\alpha_s$ ($n_f$ and $n_s$ are the
number of fermion flavors and scalars). For the $U(1)$ coupling
$\alpha_{\rm em}$ we have a similar formula as (\ref{RGE}), but in
this case
\begin{equation}\label{bU1}
b_{1}=-N_{h}\left(\frac43\,\sum Q_f^2+\frac13\,\sum Q_s^2\right)\,.
\end{equation}
In this equation, $N_h=2$ for $SU(2)_h$. For energies below $\Lambda_h^L$  we
have to replace $N_h$ in $b_1$ with $N_c=3$ (or $1$) and use the
electric charges of the quarks (leptons) rather than those of the
haplons. For the fine structure constant $\alpha_{\rm em}$ we
extrapolate its value from low energies to the Planck scale
$M_P\simeq 1.22\times 10^{19}$ GeV. At the mass of the $Z$-boson we
have $\alpha_{\rm em}^{-1}(M_Z) = 127.94 \pm 0.014$. From the mass
scale of the $Z$-boson, $\mu=M_Z$, until a scale well above
$\Lambda_h^L$, say $\mu\sim 2$ TeV, we use the RGE, taking into
account the charges of the three charged leptons and of the five
quarks, not including the top-quark:

We can follow a similar procedure to compute the QCD coupling
constant at various energies. The accurate measurement of this
constant at the $Z$ pole yields: $\alpha_s(M_Z) = 0.1184 \pm
0.0007$. At the Fermi scale $\Lambda_h^L\sim 0.3$ GeV we find
$\alpha_s(\Lambda_h^L) = 0.1010$. Well above $1$ TeV up to the
Planck scale the renormalization proceeds via haplon-pairs. The
results are summarized in Table 2.

\begin{table}[t]
\begin{center}
\begin{tabular} {|c|c|c|c|}
  \hline
  $ \phantom{x}$ & $\mu_0$ & $\mu_1$ &  $M_P$\\ \hline
  $ \phantom{x}$ & $M_Z$ & $2$ TeV & $10^{19}$ GeV  \\ \hline
  $\alpha_{\rm em}$\phantom{x} & \phantom{x}$0.007816$\phantom{x} & \phantom{x}$0.008092$\phantom{x} & \phantom{x}$0.008727$\phantom{x} \\
  $\alpha_s$\phantom{x} & $0.1184$ & $0.08187$  & $0.01370$\\
  $\alpha_h$\phantom{x} & $-$ & $0.62$  & $0.030$  \\
  \hline

\end{tabular}
\caption[]{The QED, QCD and QHD fine structure constants
$\alpha_i=g_i^2/4\pi$ at the $Z$-pole scale $\mu_0=M_Z$, at an
intermediate high energy scale $\mu_1=2$ TeV (around the haplon
continuum threshold), and at the Planck energy $M_P\sim 1.2\times
10^{19}$ GeV, for the $SU(2)_h$ chiral gauge group of QHD.}
 \end{center}
\end{table}

For the $SU(2)_h$ group, we focus here on the lefthanded sector and
assume once more $\Lambda_h^L \simeq 0.3$ TeV. Using
Eq.\,(\ref{alphasLQCD}) with $\alpha_s\to\alpha_h$ and
$\LQCD\to\Lambda_h^L$, we find e.g. $\alpha_h(2{\rm TeV}) =0.62$,
and eventually
at the Planck energy:  $\alpha_h(M_P) \simeq 0.030$.   
These results are collected in the table above, and we see that the
three couplings approach each other at the Planck scale. The details
of the unification will depend on the particular GUT group and can
be affected by Clebsch - Gordan coefficients of ${\cal O}(1)$.

If the three couplings come close at the Planck scale, interesting
consequences can be derived in connection to the time variation of
the fundamental constants, of which hints in the literature appear
quite often\,\cite{FundamentalConstants}. An exact unification is
not essential - we only need that the three couplings take fixed
values at or around $M_P$. We remark that $SU(3)\times SU(2)_L\times
SU(2)_R\times U(1)$ is a natural breakdown step for GUT groups such
as e.g. $SO(10)$. In our case we do not have spontaneous symmetry
breaking (SSB), the breaking is always meant to be dynamical. The
complete QHD group can thus be naturally linked to the GUT framework
without generating unconfined vacuum energy, in contrast to the SM.

Let us now assess a possible time change of Newton's constant $G_N$
(and hence of $M_P$). It is conceivable in the same way as one
admits a possible time change of $\alpha_{\rm
em}$\,\cite{MurphyWebbFlaumbaum2003,FundamentalConstants}. If the
QED, QCD as well as the QHD coupling constants emerge at the Planck
epoch, their primeval values should be very close and not be
time-dependent. Assuming that the Planck energy changes in time, it
implies a time evolution of the gauge couplings at lower energies,
say around the confining scale of the weak bosons, $\Lambda_h^L\sim
300$ GeV. By the same token the masses of all the particles
(including of course the baryons and the $D$-bosons) will slowly
evolve with the cosmic expansion since their binding energies are
functions of the coupling strengths. We have exemplified this
situation in (\ref{eq:LQCDH}) for a general change of particle
masses.

Let us estimate the time change of $G_N$ in the specific case of
QHD. We use the approximate time variation of $\alpha_{\rm em}$
suggested in a typical measurement where the current value of the
QED coupling is compared with that of a quasar some $12$ billion
years ago \,\cite{MurphyWebbFlaumbaum2003}: $\Delta\alpha_{\rm
em}/\alpha_{\rm em}=(-0.54\pm 0.12)\times 10^{-5}$.

From the RGE's and setting $\mu=M_P$ we can obtain the time
variation (indicated by a dot) of the Planck scale. Since
$b_1=-14/9$ in this case, we find
\begin{equation}\label{variationMPQED}
\frac{\dot{M}_P}{M_P}=-\frac{\dot{\alpha}_{\rm em}(M_Z)}{\alpha_{\rm
em}(M_Z)}\,\left[\ln\frac{M_P}{M_Z}+\frac{9\pi}{7\alpha_{\rm
em}(M_P)}\right]\,.
\end{equation}
It follows: $\Delta M_P/M_P\simeq 0.0027$ or ${\Delta G}/{G}\simeq
-0.0054$.

In a similar way we can obtain the time variation of the non-Abelian
gauge couplings $\alpha_i$ (i.e. $\alpha_s$ and $\alpha_h$) at an
arbitrary scale $\mu$ below $M_P$:
\begin{equation}\label{variationMPNonAbelian}
\frac{\dot{\alpha}_i(\mu)}{\alpha_i(\mu)}=\frac{\dot{M}_P}{M_P}\,\left[-\ln\frac{M_P}{\mu}+\frac{2\pi}{b_N\,\alpha_i(M_P)}\right]^{-1}\,,
\end{equation}
with $b_N$ defined in (\ref{RGE}).

Since $\dot{M}_P/M_P$ is fixed from (\ref{variationMPQED}), the
above equation enables us to compute the cosmic time variation of
the QCD and QHD couplings  within the last $12$ billion years at any
desired energy well above $\Lambda_h^L$, e.g. at $\mu_1=2$ TeV (cf.
Table II):
\begin{equation}\label{Delta alphai}
\frac{\Delta{\alpha}_s}{\alpha_s}\simeq 1.1\times 10^{-4}\,,\ \ \
\frac{\Delta{\alpha}_h}{\alpha_h}\simeq 6.3\,\times 10^{-4}\,.
\end{equation}
Using the definition (\ref{alphasLQCD}) for each confining scale
$\Lambda_i$\, (viz. $\Lambda_{\rm QCD},\Lambda_h^L)$ we can check
from the above formulas that their cosmic time
evolution\,\cite{FritzschSola2012} is renormalization group
invariant and is directly tied to the cosmic evolution of $M_P$
itself:
\begin{equation}\label{linkalphaLambda}
\frac{\dot{\Lambda}_i}{\Lambda_i}=\frac{\dot{\alpha}_i(\mu)}{\alpha_i(\mu)}\,\frac{2\pi}{b\,\alpha_i(\mu)}=\frac{\dot{M}_P}{M_P}\,.
\end{equation}
Numerically, $\Delta\Lambda_i/\Lambda_i\simeq 3\times 10^{-3}$ for
the indicated period.

\mysection{Conclusions} \label{sect:conclusions}

We have described theoretical models for the dark matter (DM) and
dark energy (DE) based on the idea that the basic constants of
nature are actually slowly varying functions of the cosmic
expansion, as suggested by numerous experiments.
The variation of the nuclear and particle masses, fundamental scales
and particle physics couplings (e.g. the fine structure constant and
the strong coupling of QCD) has been connected to the possible cosmic evolution of the
basic parameters $\rL$ and $G_N$ of Einstein's General Relativity
(GR).

In this framework the vacuum energy appears naturally as a dynamical quantity that varies with the cosmic expansion. If correct, we should find that as soon as the precision of the observations will improve, the so-called ``cosmological constant'' shall exhibit a mild evolution with the cosmic time and hence with the redshift. The rate of this variation will be connected to the time variation of the particle masses. In some of these models, the evolution of the gravitational coupling is also naturally involved. Thus, we expect a general dynamical feedback between the fundamental ``constants'' of the gravitational sector ($\rL(t), G_N(t),...$) and the fundamental ``constants'' of particle physics ($m_i(t), \alpha_i(t), \LQCD(t),...$), in a way fully compatible with the general covariance of the theory.

As a particular model implementation of these ideas we have considered Quantum Haplodynamics (QHD),
which is not based on the conventional SSB mechanism and it does not
lead, in contrast to the SM, to a large contribution to the
cosmological term. The DE appears here as the tiny (but observable)
dynamical change of the vacuum energy density of the expanding
background, and hence is a part of the generic response of GR to the
cosmic time variation of the masses of all the stable baryons and DM
particles in the universe.

These ideas are actually quite general and not tied to a particular model. They can be tested in future astrophysical and laboratory
tests in quantum optics, which are expected\,\cite{FritzschSola2012,FritzschSola2014}
to detect potential proton mass variations  $\lesssim10^{-14}$. While the
SM of particle physics is a successful theory, the severity of the DM and DE problems cannot be permanently hiden under the rug. Dramatically new (and testable) ideas are urgently needed!


\newpage

{\bf Acknowledgments}

It is my pleasure to thank Harald Fritzsch for the stimulating collaboration on these topics in the last few years, as well as for the smooth organization of the Int. Conf. on Flavor Physics and Mass Generation at the Nanyang Technological University (NYTU) in Singapore. I should also like to extend my thanks to the rest of the organizers, in particular to K.K. Phua, for the warm hospitality and partial support that I received from  the Institute for Advanced Study at NYTU. I am also supported in part by FPA2010-20807 (MICINN), DIUE/CUR (Generalitat de Catalunya) and CSD2007-00042
(CPAN).

\newcommand{\JHEP}[3]{ {JHEP} {#1} (#2)  {#3}}
\newcommand{\NPB}[3]{{ Nucl. Phys. } {\bf B#1} (#2)  {#3}}
\newcommand{\NPPS}[3]{{ Nucl. Phys. Proc. Supp. } {\bf #1} (#2)  {#3}}
\newcommand{\PRD}[3]{{ Phys. Rev. } {\bf D#1} (#2)   {#3}}
\newcommand{\PLB}[3]{{ Phys. Lett. } {\bf B#1} (#2)  {#3}}
\newcommand{\EPJ}[3]{{ Eur. Phys. J } {\bf C#1} (#2)  {#3}}
\newcommand{\PR}[3]{{ Phys. Rep. } {\bf #1} (#2)  {#3}}
\newcommand{\RMP}[3]{{ Rev. Mod. Phys. } {\bf #1} (#2)  {#3}}
\newcommand{\IJMP}[3]{{ Int. J. of Mod. Phys. } {\bf #1} (#2)  {#3}}
\newcommand{\PRL}[3]{{ Phys. Rev. Lett. } {\bf #1} (#2) {#3}}
\newcommand{\ZFP}[3]{{ Zeitsch. f. Physik } {\bf C#1} (#2)  {#3}}
\newcommand{\MPLA}[3]{{ Mod. Phys. Lett. } {\bf A#1} (#2) {#3}}
\newcommand{\CQG}[3]{{ Class. Quant. Grav. } {\bf #1} (#2) {#3}}
\newcommand{\JCAP}[3]{{ JCAP} {\bf#1} (#2)  {#3}}
\newcommand{\APJ}[3]{{ Astrophys. J. } {\bf #1} (#2)  {#3}}
\newcommand{\AMJ}[3]{{ Astronom. J. } {\bf #1} (#2)  {#3}}
\newcommand{\APP}[3]{{ Astropart. Phys. } {\bf #1} (#2)  {#3}}
\newcommand{\AAP}[3]{{ Astron. Astrophys. } {\bf #1} (#2)  {#3}}
\newcommand{\MNRAS}[3]{{ Mon. Not. Roy. Astron. Soc.} {\bf #1} (#2)  {#3}}
\newcommand{\JPA}[3]{{ J. Phys. A: Math. Theor.} {\bf #1} (#2)  {#3}}
\newcommand{\ProgS}[3]{{ Prog. Theor. Phys. Supp.} {\bf #1} (#2)  {#3}}
\newcommand{\APJS}[3]{{ Astrophys. J. Supl.} {\bf #1} (#2)  {#3}}

\newcommand{\Prog}[3]{{ Prog. Theor. Phys.} {\bf #1}  (#2) {#3}}
\newcommand{\IJMPA}[3]{{ Int. J. of Mod. Phys. A} {\bf #1}  {(#2)} {#3}}
\newcommand{\IJMPD}[3]{{ Int. J. of Mod. Phys. D} {\bf #1}  {(#2)} {#3}}
\newcommand{\GRG}[3]{{ Gen. Rel. Grav.} {\bf #1}  {(#2)} {#3}}




\end{document}